\documentclass[sigconf]{acmart} 
\usepackage{multirow}
\usepackage{algorithm}
\usepackage{algpseudocode}
\usepackage{epstopdf}
\usepackage{array}
\usepackage{hyperref}
\usepackage{float}
\usepackage{svg}
\usepackage{graphicx}
\usepackage{subcaption}
\usepackage{wrapfig}
\usepackage{xcolor}
\usepackage{makecell}
\usepackage{xcolor}
\usepackage{subcaption}

\usepackage[graphicx]{realboxes}
\setcopyright{acmlicensed}

\copyrightyear{2025}
\acmYear{2025}
\setcopyright{cc}
\setcctype{by}
\acmConference[ICAIF '25]{6th ACM International Conference on AI in Finance}{November 15--18, 2025}{Singapore, Singapore}
\acmBooktitle{6th ACM International Conference on AI in Finance (ICAIF '25), November 15--18, 2025, Singapore, Singapore}\acmDOI{10.1145/3768292.3770423}
\acmISBN{979-8-4007-2220-2/2025/11}

\begin{document}

\title{Return Prediction for Mean-Variance Portfolio Selection: \\How Decision-Focused Learning Shapes Forecasting Models}

\author{Junhyeong Lee}
\authornote{These authors contributed equally to this work.}
\affiliation{%
  \institution{UNIST}
  \city{Ulsan}
  \country{Republic of Korea}
}
\email{jun.lee@unist.ac.kr}

\author{Haeun Jeon}
\authornotemark[1]
\affiliation{%
  \institution{KAIST}
  \city{Daejeon}
  \country{Republic of Korea}
}
\email{haeun39@kaist.ac.kr}

\author{Hyunglip Bae}
\authornote{Corresponding author}
\affiliation{%
  \institution{KAIST}
  \city{Daejeon}
  \country{Republic of Korea}
}
\email{qogudflq@kaist.ac.kr}

\author{Yongjae Lee}
\authornotemark[2]
\affiliation{%
  \institution{UNIST}
  \city{Ulsan}
  \country{Republic of Korea}
}
\email{yongjaelee@unist.ac.kr}

\begin{abstract}

Markowitz laid the foundation of portfolio theory through mean-variance optimization (MVO). However, MVO's effectiveness depends on precise estimation of expected returns, variances, and covariances, which are typically uncertain. Machine learning models are increasingly used to estimate these parameters, trained to minimize prediction errors like MSE, which treats errors uniformly across assets. Recent studies show this leads to suboptimal decisions and propose Decision-Focused Learning (DFL), integrating prediction and optimization to improve outcomes. While studies demonstrate DFL's potential to enhance portfolio performance, the mechanisms of how DFL modifies prediction models for MVO remain unexplored. This study investigates how DFL adjusts stock return prediction models to optimize MVO decisions. We show that DFL's gradient tilts MSE-based prediction errors by the inverse covariance matrix $\Sigma^{-1}$, incorporating inter-asset correlations into learning, while MSE treats each asset independently. This tilting creates systematic biases where DFL overestimates returns for portfolio assets while underestimating excluded assets. Our findings reveal why DFL achieves superior performance despite higher prediction errors. The strategic biases are features, not flaws.

\end{abstract}



\renewcommand{\shortauthors}{Lee et al.}

\maketitle
\section{Introduction}
\label{sec:intro}

Decision-making is a crucial aspect across various fields, where optimization is often employed to guide us toward the best possible outcome quantitatively. In ideal scenarios, where the values of all parameters in the optimization formula are known, a mathematically optimal solution can be determined with precision. However, in most real-world situations, parameters are often uncertain. The quality of the decisions derived from optimization heavily depends on how accurately these uncertain parameters are estimated. Therefore, it is essential to ensure accurate estimation of these parameters to achieve high-quality decision-making outcomes.

This is particularly evident in the field of portfolio optimization, where financial decisions are made under uncertainty. Harry Markowitz laid the foundation of portfolio theory based on mean-variance optimization (MVO) \cite{e5a1bb8f-41b7-35c6-95cd-8b366d3e99bc}. The objective of MVO is to construct an investment portfolio that maximizes return for a given level of risk or minimizes risk for a given level of expected return. While the MVO has been fundamental to investment management \cite{kim2021mean}, its effectiveness depends on the accurate estimates of the expected return, variance, and covariance of asset returns, which are often uncertain in practice. Despite this uncertainty, decision makers who utilize MVO should have their own estimates of expected returns and risks of financial assets. Thus, many researchers and practitioners have questioned how the input parameters of MVO should be estimated. To this, Markowitz is said to have responded with wit and grace, \textit{``That's your job, not mine.''} \cite{sexauer2024harry}

In relation to \textit{``your job''}, many studies have been conducted to investigate the impact of estimation errors in input parameters on MVO. Markowitz showed that MVO can maximize the effect of input parameter estimation errors, which can lead to inferior results compared to an equally weighted portfolio \cite{michaud1989markowitz}. Other studies have discussed the importance of input parameter settings in the MVO framework \cite{best1991sensitivity, best1991sensitivityms, kallberg1984mis}. In addition, some researchers have analyzed how the MVO optimal portfolio or the distribution of all possible portfolios change as the input parameters change \cite{chopra1993effect, kallberg1984mis, best1991sensitivity, best1991sensitivityms, chung2022effects, pan2024bpqp}. While these studies reveal how the degree of estimation errors affects the MVO framework, they do not go further into how the shape of estimation errors affects the MVO framework.

In practice, machine learning has become very useful in the estimation of parameters, and decisions are made through optimization based on the machine learning estimates as inputs \cite{lee2023overview}. Hence, the so-called \textit{Predict-then-Optimize} method can be seen as a two-stage method. The prediction and optimization stages are separated, and thus, the prediction stage is solely concerned with enhancing prediction accuracy, such as the mean squared error (MSE). Recent studies have argued that a prediction model that minimizes traditional prediction losses, such as MSE, may not be optimal for decision-making in the subsequent optimization stage. To overcome this issue, a framework called \textit{Decision-Focused Learning (DFL)} has been proposed \cite{donti2017task, elmachtoub2022smart, wilder2019melding, poganvcic2020differentiation, mandi2022decision, shah2022decision}. DFL has been studied in various fields, and portfolio optimization is no exception. A couple of studies \cite{butler2023integrating, costa2023distributionally, anis2025end} have shown that DFL can be implemented for portfolio optimization and can enhance investment performance. However, they have not analyzed the detailed characteristics of the DFL prediction model. 

We discovered an interesting phenomenon when applying DFL to MVO. As shown in Table~\ref{tab:max_weight_intro}, while MSE-trained models diversify portfolios more as risk aversion $\lambda$ increases, DFL-trained models exhibit the opposite behavior by consistently concentrating portfolios to the extreme, using only the minimum number of assets allowed by constraints, regardless of $\lambda$. This suggests DFL identifies extreme concentration as optimal for MVO performance.

\begin{table}[t]
\centering
\caption{Average number of assets in portfolios. Rows show maximum weight constraints ($k$ where each weight $\leq 1/k$), columns show risk aversion levels ($\lambda$). Values are mean$\pm$std over 5 seeds. DFL consistently uses exactly $k$ assets while MSE uses more, showing DFL's extreme concentration.}
\label{tab:max_weight_intro}
\vspace{0.3em}
\footnotesize
\renewcommand{\arraystretch}{1.15}
\begin{tabular*}{\columnwidth}{@{\extracolsep{\fill}}clccccc@{}}
\toprule
& & \multicolumn{4}{c}{\textbf{Number of Active Assets}} \\
\cmidrule(lr){3-6}
\textbf{Max weight} $\leq \mathbf{1/k}$ & \textbf{Loss} & \boldmath$\lambda$\textbf{=0.1} & \boldmath$\lambda$\textbf{=0.5} & \boldmath$\lambda$\textbf{=1.0} & \boldmath$\lambda$\textbf{=5.0} \\
\midrule
\multirow{2}{*}{$k=1$} 
 & MSE & 1.4±0.0 & 1.9±0.0 & 2.3±0.1 & 4.0±0.3 \\
 & DFL & 1.0±0.0 & 1.0±0.0 & 1.0±0.0 & 1.0±0.0 \\
\midrule
\multirow{2}{*}{$k=5$}
 & MSE & 6.1±0.1 & 6.4±0.1 & 6.7±0.2 & 7.5±0.1 \\
 & DFL & 5.0±0.0 & 5.0±0.0 & 5.0±0.0 & 5.0±0.0 \\
\midrule
\multirow{2}{*}{$k=10$}
 & MSE & 11.1±0.1 & 11.3±0.1 & 11.4±0.1 & 11.8±0.1 \\
 & DFL & 10.0±0.0 & 10.0±0.0 & 10.0±0.0 & 10.0±0.0 \\
\bottomrule
\end{tabular*}
\vspace{-12.5pt}
\end{table}

The objective of this study is to gain insight into how the prediction model changes when DFL is applied to MVO. To be more specific, we wish to answer the following question:

\noindent \textit{``How does DFL modify the stock return prediction model to produce optimal decisions in MVO? MSE treats the errors of all assets equally, but how does DFL reduce the errors of different assets differently?''}

\noindent Note that this question has not been answered in the two streams of research, estimation errors in MVO and DFL in MVO. A systematic approach to this question would enable us to understand how we should predict stock returns when we are going to use them to construct efficient portfolios.

\begin{figure*}[t]
    \centering
    \includegraphics[width=18cm]{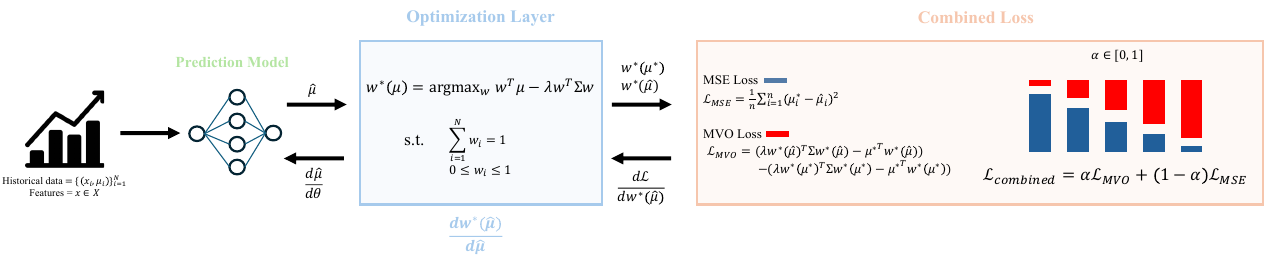}
    \caption{DFL training procedure for MVO. The prediction model outputs $\hat{\mu}$, which determines portfolio weights $w^*(\hat{\mu})$ through the optimization layer. The combined loss $\mathcal{L}_{combined} = \alpha\mathcal{L}_{MVO} + (1-\alpha)\mathcal{L}_{MSE}$ balances prediction accuracy and decision quality. The optimization layer enables computation of $\frac{dw^*(\hat{\mu})}{d\hat{\mu}}$, which is necessary for calculating the MVO loss gradient.}
    \label{fig:Figure1}
\end{figure*}

\section{Background}
\label{sec:backround}

\subsection{Decision-Focused Learning (DFL)}
\label{sec:dfl}

\begin{equation}
\begin{aligned}
w^*(c) &= \arg \min_w f(w,c) \\
& \quad \text{s.t.} \quad g(w,c) \leq 0 \\
& \quad \phantom{\text{s.t.}} \quad h(w,c) = 0
\end{aligned}
\label{eq:eq1}
\end{equation}

\noindent
\textbf{Predict-then-Optimize.} Following the notations of \cite{mandi2023decision}, the general objective of an optimization problem is to find a solution $w^*(c)$, where $w$ represents the decision variables and $c$ the parameters. The solution $w^*(c)$ minimizes the objective function $f(w,c)$ while satisfying the inequality constraint $g(w,c) \le 0$ and equality constraint $h(w,c)=0$. The so-called "Predict-then-Optimize" framework proceeds with prediction before optimization, also known as two-stage learning. Initially, a machine learning model $F_{\theta}$ generates $\hat{c} = F_{\theta}(x)$, with $\theta$ being the model parameters and $x$ the input features for predicting $\hat{c}$. Subsequently, optimization is performed using the predicted parameter $\hat{c}$. In this case, traditional ML training methods are used to predict the ground truth $c^*$ accurately. Commonly, MSE or cross-entropy losses are used to minimize the difference between the ground truth $c^*$ and the predicted parameter $\hat{c}$, thus training the model.

\noindent
\textbf{Decision-Focused Learning.} Many studies (e.g., \cite{donti2017task, elmachtoub2022smart, wilder2019melding}) suggested that the predict-then-optimize framework often results in suboptimal outcomes, because the prediction and optimization stages are separated. Minimizing prediction errors measured by MSE or cross-entropy loss is not necessarily beneficial to the subsequent decision-making stage. To overcome this limitation, the DFL framework has been proposed, which can be seen as a new model training methodologies that consider both prediction and optimization stages holistically. 

\begin{equation}
\begin{aligned}
Regret(w^*(\hat{c}),c^*) = f(w^*(\hat{c}),c^*) - f(w^*(c^*),c^*)
\end{aligned}
\label{eq:regret}
\end{equation}

In DFL, a machine learning model is trained to minimize a loss function that reduces the decision-making error when the actual decision is realized through $w^*(\hat{c})$. Specifically, $Regret(w^*(\hat{c}), c)$, which measures the suboptimality of the decision made via $w^*(\hat{c})$, is considered in most cases. The prediction model is trained to predict $\hat{c}$, which is most helpful for optimal decision making.

While the concept of DFL is straightforward, there are some obstacles when implementing it. The major issue in DFL implementation is the difficulty of calculating gradients for model training. Let $\mathcal{L}$ be the DFL loss, analogous to the concept of regret. In order to proceed with gradient-based learning, it is necessary to differentiate $\mathcal{L}$ with respect to the model parameter $\theta$. The gradient can be expressed as follows based on the chain rule:

\begin{equation}
\begin{aligned}
\frac{d \mathcal{L}(w^*(\hat{c}),c^*)}{d \theta} = \frac{d \mathcal{L}(w^*(\hat{c}),c^*)}{d w^*(\hat{c})} \frac{d w^*(\hat{c})}{d \hat{c}} \frac{d \hat{c}}{d \theta}
\end{aligned}
\label{eq:eq3}
\end{equation}

The first term on the right-hand side should not be a problem, because the DFL loss consists of the objective function of the optimization problem, and thus, it should be mostly differentiated with respect to $w^*$ analytically. The third term can be computed in the same way as for the usual gradient-based learning of most prediction models. However, the second term is the gradient of the optimal solution of an optimization problem, which is extremely tricky. Even if the solution is continuous, the second gradient must be calculated through an argmin or argmax operation \cite{wilder2019melding}.

Several approaches have been developed to integrate optimization problems into neural networks \cite{amos2017optnet, agrawal2019differentiable}. To avoid direct calculation of the second gradient, recent work has introduced surrogate functions \cite{elmachtoub2022smart, mulamba2020contrastive, mandi2022decision, shah2022decision, shah2024leaving, jeon2025locally}.
Note that this study aims to analyze how DFL affects the prediction model, and thus, the efficiency of training is not an important issue. Hence, we calculate loss function values by directly solving optimization problems.




\subsection{Mean-Variance Optimization (MVO)}

\noindent
\textbf{Model Formulation.} A typical formulation for mean-variance optimization developed by Markowitz \cite{e5a1bb8f-41b7-35c6-95cd-8b366d3e99bc} is as follows:

\begin{equation}
\begin{aligned}
w^*({\mu}) &= \arg \max_w \ w^T{\mu} - \lambda w^T \Sigma w \\
& \quad \text{s.t.} \quad \sum_{i=1}^N w_i = 1, \\
& \quad \phantom{\text{s.t.} \quad} 0 \leq w_i \leq 1 \quad \text{for } i = 1, \ldots, N
\end{aligned}
\label{eq:eq4}
\end{equation}

Here, $w$ represents the portfolio weights of the $N$ risky assets, which are constrained to have a sum equal to 1 and to be between 0 and 1. $\mu$ is the expected return of the assets, $\lambda$ is the risk aversion, and $\Sigma$ is the covariance of asset returns. This optimization problem allows us to maximize the portfolio returns, $w^T \mu$, while considering a risk penalty, $\lambda w^T \Sigma w$. Note that the optimal portfolio weight $w^*$ is represented as a function of expected return $\mu$, because the focus of this study is on how the optimal portfolio $w^*$ changes with respect to the prediction of $\mu$.

As mentioned in \hyperref[sec:intro]{Section 1}, estimating uncertain parameters for optimization is challenging, as financial asset returns are highly volatile. To address this, various approaches have been proposed, including robust optimization~\cite{kim2023robustness}, Black-Litterman model~\cite{black1992global}, Bayesian approach~\cite{jorion1986bayes}, and risk factor models~\cite{ross1976arbitrage}.

\noindent
\textbf{Estimation of Returns and Covariances.} As mentioned in Section \ref{sec:intro}, there has been extensive research on the estimation errors of the input parameter of the MVO framework (i.e., $\mu$ and $\Sigma$). Early work highlighted the relative importance of estimation errors in MVO parameters \cite{chopra1993effect}. Simple perturbations on mean, variance, and covariance revealed that errors in $\mu$ have a relatively greater impact on the optimal objective compared to errors in $\Sigma$. The sensitivity of optimal portfolios to estimation errors in $\mu$ and $\Sigma$ has been quantified theoretically, showing that the relative impact of covariance matrix errors mainly depends on the Sharpe ratio \cite{palczewski2014theoretical}. Additional research has been conducted employing different constraints, problem formulations, or assumptions \cite{kan2007optimal, kaut2007stability}. This makes it challenging to draw general conclusions. To address this issue, recent work has analyzed how estimation errors affect the Sharpe ratio distribution of all possible portfolios, which can be analytically calculated using previously developed methods \cite{chung2022effects, kim2016uniformly}. These findings indicate that correlations play a more significant role in the MVO framework. While these studies reveal how the degree of estimation errors affects the MVO framework, they do not go further into how the shape of estimation errors affects the MVO framework.

\vspace{-4pt}
\section{Mechanism of DFL for MVO}
\label{sec:Section3}

\begin{table*}[t]
\vspace{-1em}
\caption{Portfolio performance metrics for varying $\lambda$ and $\alpha$. Models with $\alpha > 0$ consistently outperform pure MSE ($\alpha = 0$), with optimal performance typically at intermediate values ($\alpha \in [0.25, 0.75]$). DFL improves both returns and risk metrics, achieving higher Sharpe ratios and often lower maximum drawdowns. Bold values indicate best performance for each metrics.}
\label{tab:Table1}
\vspace{0.5em}
\centering
\small
\setlength{\tabcolsep}{2.5pt}
\renewcommand{\arraystretch}{0.95}
\begin{tabular}{@{}lccccc@{\hspace{1.2em}}lccccc@{}}
\toprule
\multicolumn{6}{c}{\textbf{Panel A. DOW 30 Dataset}} & \multicolumn{6}{c}{\textbf{Panel B. S\&P 100 Dataset}} \\
\cmidrule(r){1-6} \cmidrule(l){7-12}
$\lambda$ & $\alpha$ & Return ($\uparrow$) & Sharpe ($\uparrow$) & MDD ($\downarrow$) & Wealth ($\uparrow$) & $\lambda$ & $\alpha$ & Return ($\uparrow$) & Sharpe ($\uparrow$) & MDD ($\downarrow$) & Wealth ($\uparrow$) \\
\midrule
\multirow{5}{*}{0.1} 
& 0.00 & 0.181 & 0.681 & 0.253 & 1.459 &
\multirow{5}{*}{0.1} 
& 0.00 & 0.209 & 0.734 & 0.266 & 1.184 \\
& 0.25 & 0.323 & 1.138 & 0.230 & 1.669 &
& 0.25 & 0.256 & 0.744 & 0.327 & 1.673 \\
& 0.50 & \textbf{0.398} & \textbf{1.228} & 0.261 & \textbf{2.138} &
& 0.50 & 0.134 & 0.513 & 0.278 & 1.239 \\
& 0.75 & 0.177 & 0.699 & 0.268 & 1.408 &
& 0.75 & 0.091 & 0.386 & 0.319 & 1.277 \\
& 1.00 & 0.143 & 0.593 & \textbf{0.226} & 1.275 &
& 1.00 & \textbf{0.292} & \textbf{1.179} & \textbf{0.249} & \textbf{1.701} \\
\midrule
\multirow{5}{*}{0.5} 
& 0.00 & 0.125 & 0.529 & 0.251 & 1.250 &
\multirow{5}{*}{0.5} 
& 0.00 & 0.153 & 0.578 & 0.276 & 1.126 \\
& 0.25 & 0.192 & 0.754 & 0.258 & 1.459 &
& 0.25 & 0.187 & 0.769 & 0.295 & 1.339 \\
& 0.50 & \textbf{0.347} & \textbf{1.302} & \textbf{0.211} & \textbf{1.813} &
& 0.50 & 0.231 & 0.701 & 0.295 & 1.680 \\
& 0.75 & 0.248 & 1.051 & 0.230 & 1.520 &
& 0.75 & \textbf{0.303} & \textbf{1.217} & \textbf{0.230} & \textbf{1.763} \\
& 1.00 & 0.151 & 0.641 & 0.268 & 1.354 &
& 1.00 & 0.183 & 0.769 & 0.267 & 1.512 \\
\midrule
\multirow{5}{*}{1.0} 
& 0.00 & 0.115 & 0.519 & 0.242 & 1.234 &
\multirow{5}{*}{1.0} 
& 0.00 & 0.129 & 0.540 & 0.242 & 1.090 \\
& 0.25 & 0.165 & 0.775 & 0.221 & 1.383 &
& 0.25 & 0.237 & 0.746 & 0.267 & 1.772 \\
& 0.50 & \textbf{0.327} & 1.150 & 0.218 & \textbf{1.669} &
& 0.50 & \textbf{0.319} & \textbf{1.260} & \textbf{0.214} & \textbf{1.773} \\
& 0.75 & 0.312 & \textbf{1.311} & \textbf{0.203} & 1.651 &
& 0.75 & 0.201 & 0.801 & 0.237 & 1.410 \\
& 1.00 & 0.180 & 0.763 & 0.256 & 1.397 &
& 1.00 & 0.252 & 0.949 & 0.287 & 1.569 \\
\midrule
\multirow{5}{*}{5.0} 
& 0.00 & 0.091 & 0.531 & \textbf{0.195} & 1.163 &
\multirow{5}{*}{5.0} 
& 0.00 & 0.144 & 0.838 & \textbf{0.194} & 1.258 \\
& 0.25 & \textbf{0.217} & \textbf{0.932} & 0.203 & \textbf{1.566} &
& 0.25 & 0.203 & 0.831 & 0.250 & 1.523 \\
& 0.50 & 0.187 & 0.756 & 0.215 & 1.473 &
& 0.50 & 0.136 & 0.585 & 0.219 & 1.347 \\
& 0.75 & 0.197 & 0.871 & 0.211 & 1.379 &
& 0.75 & 0.193 & 0.855 & 0.237 & 1.464 \\
& 1.00 & 0.164 & 0.738 & 0.223 & 1.361 &
& 1.00 & \textbf{0.213} & \textbf{0.928} & 0.276 & \textbf{1.611} \\
\bottomrule
\end{tabular}
\vspace{-1em}
\end{table*}

In this section, we seek to provide a theoretical explanation for why such prediction bias is amplified under DFL in asset price prediction. Specifically, we analyze the theoretical differences between the MVO loss and the MSE loss in the context of the unconstrained MVO problem, where a closed-form solution exists. This analysis aims to deliver insights into the underlying mechanisms that cause DFL to induce distinct patterns of prediction bias, particularly in comparison to traditional prediction-focused learning.

The goal of MVO is to find a portfolio $w^\ast$ that maximizes the return adjusted for risk, which can be expressed through an optimization problem:
\begin{equation}
w^*({\mu}) = \arg \max_w \ w^T{\mu} - \lambda w^T \Sigma w
\label{mvo_utility}
\end{equation}

Here, $\mu$ is the vector of expected returns, $\Sigma$ is the covariance matrix, and $\lambda$ is the risk aversion coefficient. For the unconstrained case, there is an analytical solution for the optimal portfolio $w^*$. Based on a prediction of expected returns $\hat{\mu}$, the solution is as follows:

\begin{equation}
w^*({\hat{\mu}}) = \arg \max_w \left( w^T \hat{\mu} - \lambda w^T \Sigma w \right) = \frac{1}{2\lambda} \Sigma^{-1} \hat{\mu}
\label{mvo_max}
\end{equation}

Based on this expression, the utility $U(\mu^\ast, \hat{\mu})$ of a portfolio $w^*(\hat{\mu})$, when evaluated using the ground truth expected return $\mu^*$, is as follows:

\begin{equation}
\begin{aligned}
U(\mu^*, \hat{\mu}) &= w^*(\hat{\mu})^T \mu^* - \lambda w^*(\hat{\mu})^T\Sigma w^*(\hat{\mu}) \\
&= \frac{1}{2\lambda} \hat{\mu}^T \Sigma^{-1} \mu^* - \frac{1}{4\lambda} \hat{\mu}^T \Sigma^{-1} \hat{\mu}
\end{aligned}
\label{mvo_mu}
\end{equation}

Then, a DFL model is trained using the gradient of the realized utility $U(\mu^*, \hat{\mu})$ with respect to the model prediction $\hat{\mu}$, which can be calculated as:

\begin{equation}
\frac{\partial U(\mu^*, \hat{\mu})}{\partial \hat{\mu}} = \frac{1}{2\lambda} \Sigma^{-1} (\mu^* - \hat{\mu})
\label{mvo_prime}
\end{equation}

The $i$-th element of the gradient is:

\begin{equation}
\frac{\partial U(\mu^*, \hat{\mu})}{\partial \hat{\mu}_i} = {\color{lightgray}\frac{1}{2\lambda}} \textcolor{red}{\Sigma^{-1}_i} (\mu^* - \hat{\mu})
\label{mvo_i}
\end{equation}

Here, $\Sigma^{-1}_i$ denotes the $i$th row of $\Sigma^{-1}$. The gradient of MSE with respect to $\hat{\mu}_i$ is:

\begin{equation}
\frac{\partial \text{MSE}(\mu^*, \hat{\mu})}{\partial \hat{\mu}_i} \propto (\hat{\mu}_i - \mu^*_i)
\label{mse}
\end{equation}

Hence, the gradient of the DFL model $\frac{1}{2\lambda} \Sigma^{-1}_i (\mu^* - \hat{\mu})$ can be interpreted as tilting the MSE-based prediction error vector $(\mu^* - \hat{\mu})$ by the inverse covariance matrix $\Sigma^{-1}$. While MSE treats each asset's prediction error independently as in Eq. \eqref{mse}, DFL transforms these errors through $\Sigma^{-1}$ as shown in Eq. \eqref{mvo_i}, incorporating inter-asset correlations into the learning process. We hypothesize that \emph{this covariance-weighted mechanism leads DFL to strategically modify asset-specific predictions}, ultimately creating the heavily concentrated portfolios observed in Table 1.
\section{Experiment}
\label{sec:Section4}


\begin{figure*}[ht!]
\centering
\begin{subfigure}{0.6\textwidth}
    \centering
    \includegraphics[width=\textwidth, height=1.6in, keepaspectratio]{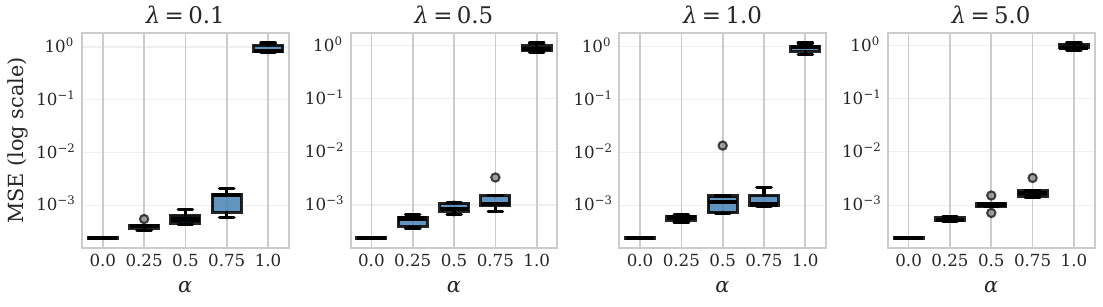}
    \caption{MSE loss (log scale)}
\end{subfigure}\\[-0.1em]
\begin{subfigure}{0.6\textwidth}
    \centering
    \includegraphics[width=\textwidth, height=1.6in, keepaspectratio]{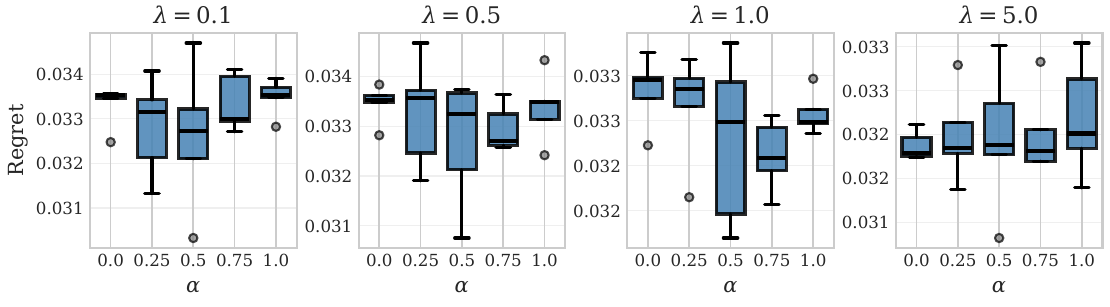}
    \caption{Regret}
\end{subfigure}
\vspace{-0.3em}
\caption{MSE and regret losses on DOW30 test set for varying $\alpha$. The boxes show distribution across 5 random seeds. MSE increases exponentially while regret decreases as $\alpha$ increases, showing DFL's trade-off between prediction accuracy and portfolio performance. The regret improvement diminishes at $\lambda = 5.0$ where risk penalty dominates the optimization.}
\label{fig:mse_regret}
\end{figure*}

\begin{figure}[t!]
    \centering
    \includegraphics[width=0.9\columnwidth]{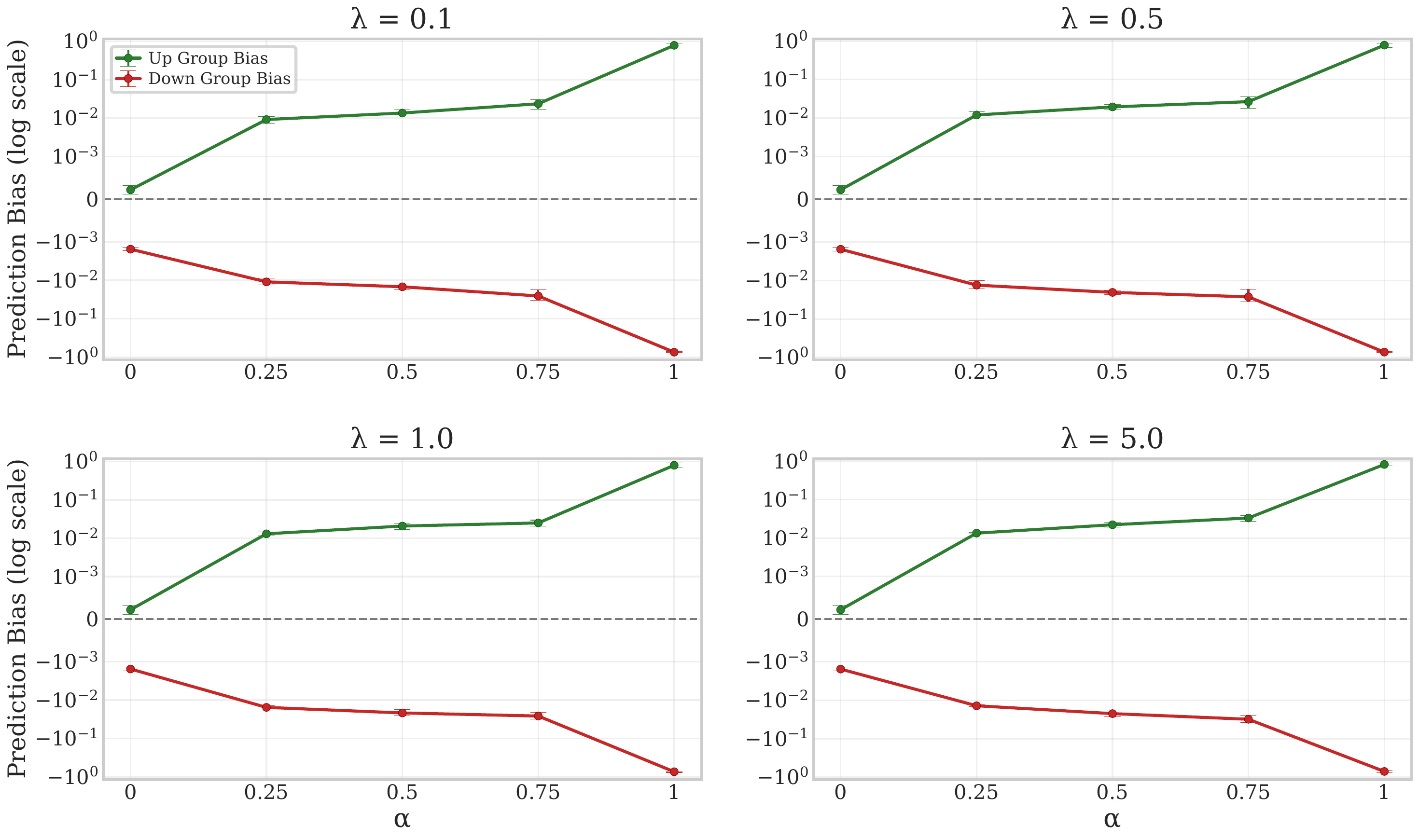}
    \caption{Prediction bias across Up/Down assets in DOW 30. As $\alpha$ increases, the Up group becomes increasingly overestimated while the Down group becomes underestimated, reaching extreme polarization at $\alpha = 1$.}
    \vspace{-10pt}
\end{figure}

\subsection{Dataset}

For the dataset, we used the DOW30 and S\&P100 asset universes, covering the period from 01-Jan-2010 to 31-Dec-2024. These asset universes are widely used in various academic studies, and by employing universes of different sizes, we aimed to demonstrate that DFL models offer a decision-making advantage over models trained with traditional MSE loss. For the prediction model, daily returns over a 60-day look-back period were used as input features $X$. Prediction models were trained to predict the returns of $N$ risky assets at time $t+1$ based on the returns of the past 60 days $X$. For the MVO problem, two parameters, expected return and covariance matrix, are required. For the expected return, we train a prediction model to obtain $\hat{\mu}$. For the covariance matrix, historical covariances were calculated based on the same look-back period. Both the prediction of expected returns and the covariances are calculated on a daily rolling basis.

\subsection{Loss Functions}

\textbf{DFL Loss for MVO.} In this study, we focus on how the prediction of expected returns $\mu$ changes when DFL is implemented for MVO. That is, for covariances, we use the sample covariance matrix $\hat{\Sigma}$ calculated using historical data. We define the DFL loss for MVO based on the regret loss defined in Eq. \ref{eq:regret} as follows:
\begin{equation}
\begin{aligned}
\mathcal{L}_{\text{MVO}} &= Regret(w^*(\hat{\mu}),\mu^*) 
\\ &= f(w^*(\hat{\mu}),\mu^*) - f(w^*(\mu^*),\mu^*) 
\\ &= (\lambda w^*(\hat{\mu})^T \Sigma w^*(\hat{\mu}) - {\mu^*}^T w^*(\hat{\mu})) 
\\ &\quad - (\lambda w^*(\mu^*)^T \Sigma w^*(\mu^*) - {\mu^*}^T w^*(\mu^*))
\end{aligned}
\label{eq:mvoloss}
\end{equation}
A machine learning model $F_{\theta}$ generates $\hat{\mu}$, the `predicted' expected return. The goal is to train the model $F_{\theta}$ in such a way that it minimizes the difference between the objective value obtained with the prediction $\hat{\mu}$ and the objective value obtained with the ground truth $\mu^*$.

\noindent \textbf{Combined Loss.} We do not simply compare a prediction model trained to minimize prediction error and a prediction model trained with DFL. Instead, we analyze the changes in the prediction model as it gradually becomes more decision-focused. In this regard, we define a combined loss as the weighted sum of MVO loss Eq. \ref{eq:mvoloss} and mean squared error (MSE), which is the most common loss function for prediction models.
\begin{equation}
\begin{aligned}
\mathcal{L}_{\text{Combined}} = \alpha \mathcal{L}_{\text{MVO}} + (1-\alpha) \mathcal{L}_{\text{MSE}}
\end{aligned}
\end{equation}
In the equation above, $\alpha$ is a constant between 0 and 1, which controls the balance between MVO and MSE. As $\alpha$ increases, more weight is given to the MVO loss, and as $\alpha$ decreases, more weight is assigned to the MSE loss. Note that MSE is defined as $\frac{1}{N} \sum_{i=1}^N (\mu^*_i - \hat{\mu}_i)^2$. For numerical experiments, the MSE loss may be multiplied by a positive scalar to match the scale of MVO loss.

\subsection{Experimental Setup}

In this study, our goal is not only to verify that DFL outperforms traditional machine learning methods, but also to demonstrate how DFL improves decision quality compared to standard MSE training and reshapes the predictive model to better align with downstream optimization. The training process is illustrated in Figure~\ref{fig:Figure1}. A prediction model outputs $\hat{\mu}$, which is used to compute the optimal portfolio $w^*(\hat{\mu})$ by solving Eq.~\eqref{eq:eq4}. The resulting decision is compared with the ground truth returns $\mu^*$ and decision $w^*(\mu^*)$, and the combined loss is backpropagated through the optimization layer, enabling gradient flow through the decision-making process.

We adopt a 4-layer MLP with $60 \times N$ input features and train it for up to 200 epochs with early stopping. Models are optimized using AdamW with a learning rate of $10^{-4}$ and a batch size of 16. We evaluate models with $\alpha \in \{0, 0.25, 0.5, 0.75, 1.0\}$ and $\lambda \in \{0.1, 0.5, 1.0, 5.0\}$ to systematically analyze how the trade-off between prediction accuracy and decision quality changes across risk preferences. Each experiment is repeated five times with different random seeds, and mean and standard deviation are reported in all tables and figures. Most model hyperparameter settings follow prior DFL studies~\cite{shah2022decision,pan2024bpqp} and time series studies~\cite{wu2022timesnet,zeng2023transformers}.

\begin{figure*}[t!]
    \centering
    \begin{subfigure}[b]{0.48\textwidth}
        \centering
        \includegraphics[width=\textwidth]{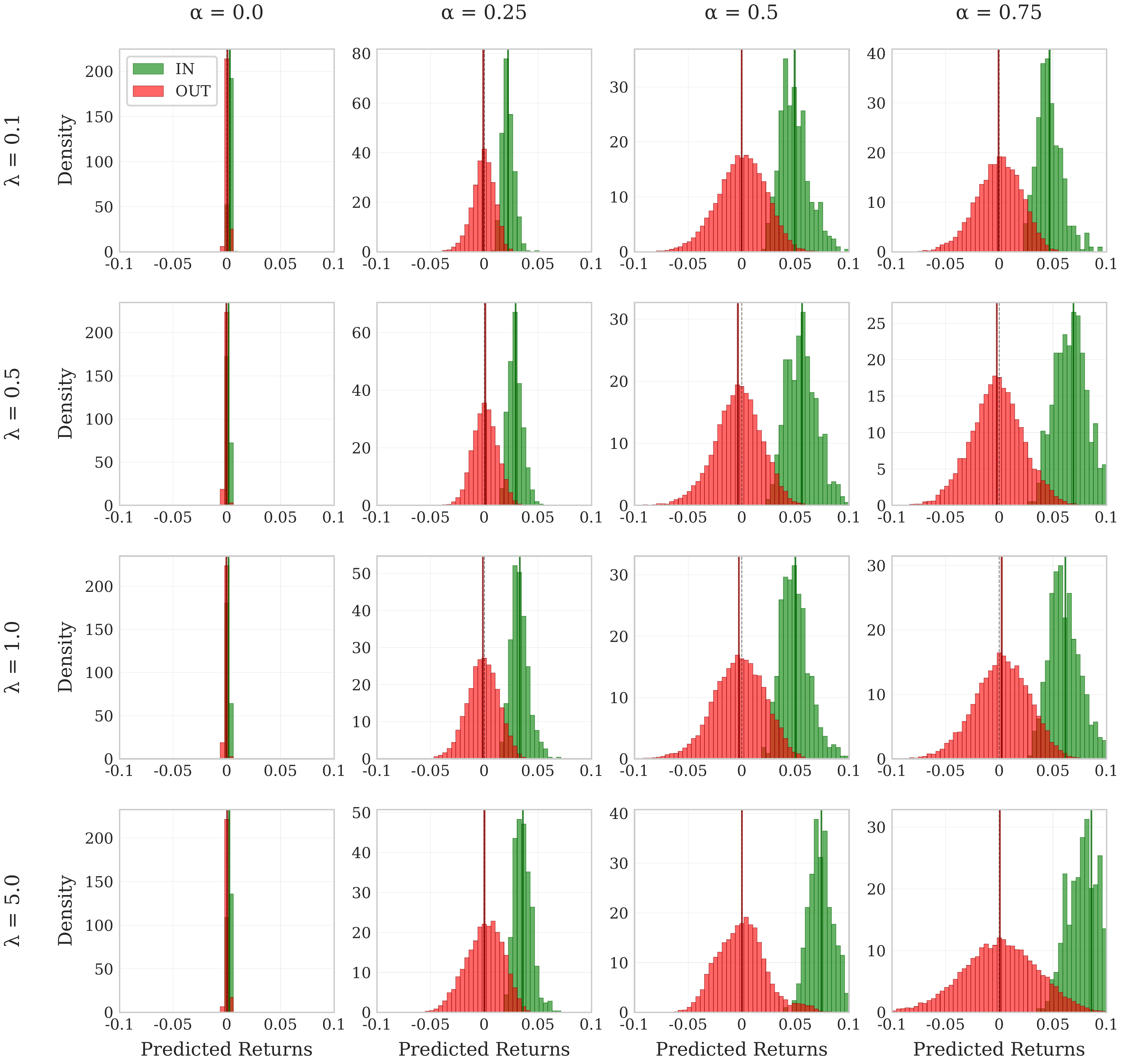}
        \caption{DOW30}
        \label{fig:distributions_dow30}
    \end{subfigure}
    \hfill
    \begin{subfigure}[b]{0.48\textwidth}
        \centering
        \includegraphics[width=\textwidth]{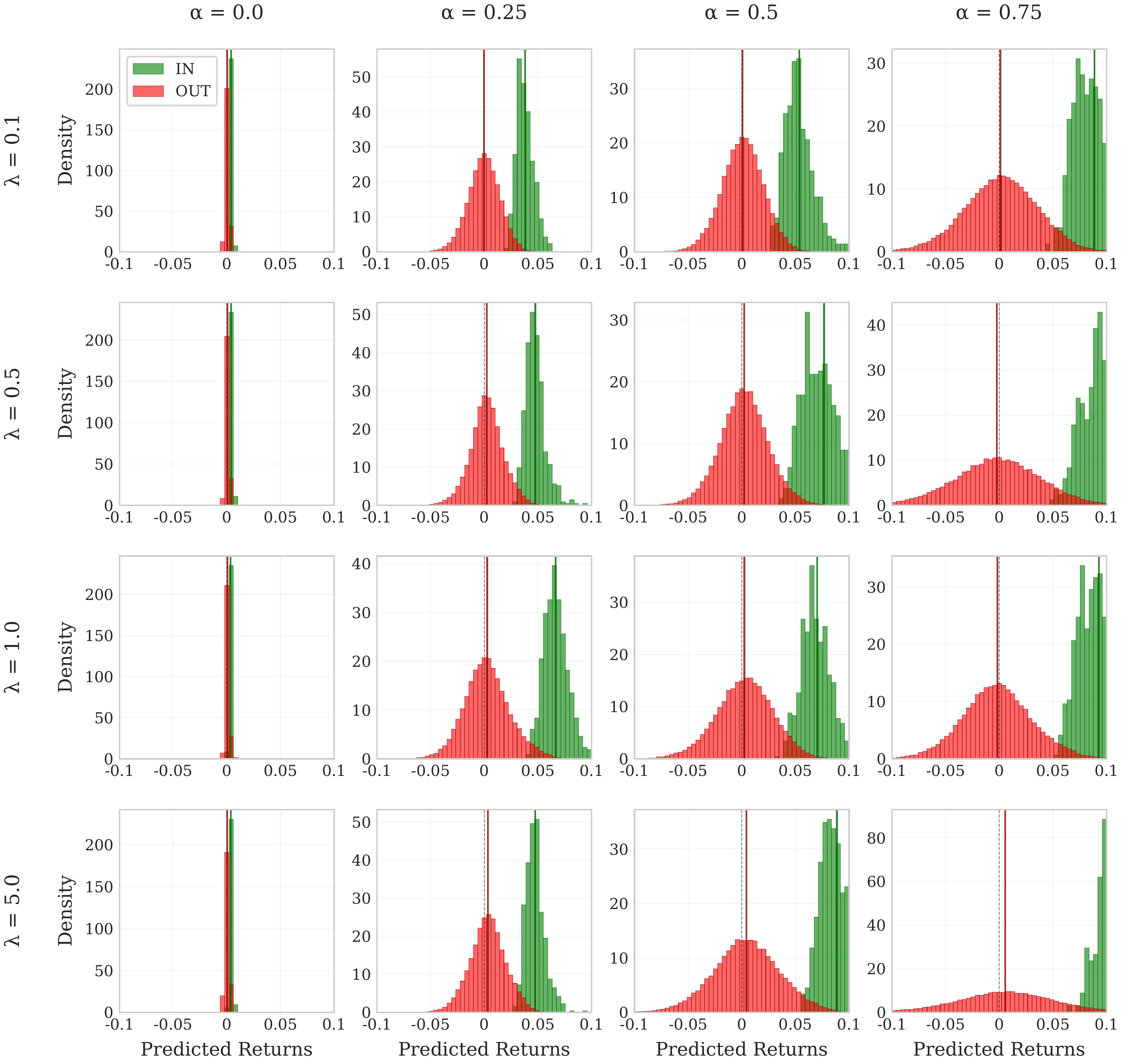}
        \caption{S\&P100}
        \label{fig:distributions_sp100}
    \end{subfigure}
    \caption{Predicted return distributions for IN/OUT portfolio groups across different $\lambda$ and $\alpha$ values. As $\alpha$ increases, the separation between IN and OUT group distributions widens. The case of $\alpha = 1$ is excluded due to extreme distribution separation.}
    \label{fig:predicted_return_distributions}
\end{figure*}
\section{Experimental Results}
\label{section5}

In Section ~\ref{sec:Section3}, we established that DFL in MVO introduces $\Sigma^{-1}$ as a tilting factor that influences asset specific predictions. Building on this theoretical foundation, we now empirically investigate how this mechanism manifests in practice, specifically examining whether and how it leads to systematic overestimation or underestimation of asset returns. Our analysis proceeds in two stages. First, we verify that our models trained with DFL indeed achieve superior decision quality compared to baseline MSE models, confirming the effectiveness of decision focused learning. Second, and more importantly, we analyze the prediction bias patterns in DFL models to empirically validate our hypothesis about how $\Sigma^{-1}$ shapes the learning process.

It is crucial to emphasize that the primary objective of our experiments is not merely to demonstrate performance improvements through DFL, which has been well established in prior work. Rather, we aim to uncover the underlying mechanisms through which DFL operates by systematically analyzing how it transforms prediction patterns. This mechanistic understanding provides insights into why decision focused learning succeeds and how it fundamentally differs from traditional prediction focused approaches.

\subsection{Model Performance}

Table~\ref{tab:Table1} reports annualized return, Sharpe ratio, maximum drawdown (MDD), and final portfolio wealth for both datasets. Our results demonstrate that intermediate $\alpha$ values consistently outperform both extremes. $\alpha = 0$ exhibit the poorest risk-adjusted returns, suggesting that accurate return predictions alone do not translate to optimal portfolio decisions. $\alpha = 1.0$ achieves better performance but remains suboptimal. $\alpha \in [0.25, 0.75]$ achieve superior Sharpe ratios while maintaining reasonable drawdowns. This pattern holds across both DOW~30 and S\&P~100 datasets, indicating robustness to different asset universes. The risk aversion parameter $\lambda$ exhibits expected behavior across all models. Increasing $\lambda$ reduces portfolio drawdowns and volatility, with the most pronounced effects at $\lambda = 5.0$. However, the optimal risk-return balance varies with $\alpha$, suggesting that decision-focused components alter how models respond to risk constraints. Importantly, DFL significantly reduces portfolio regret on the test set. As shown in Figure~\ref{fig:mse_regret}, models with $\alpha > 0$ consistently achieve lower test regret compared to $\alpha = 0$.

\begin{figure*}[t!]
    \centering
    \subfloat[$\mathbf{\alpha=0}$\label{fig:alpha0_subfig}]{%
        \includegraphics[width=0.49\textwidth]{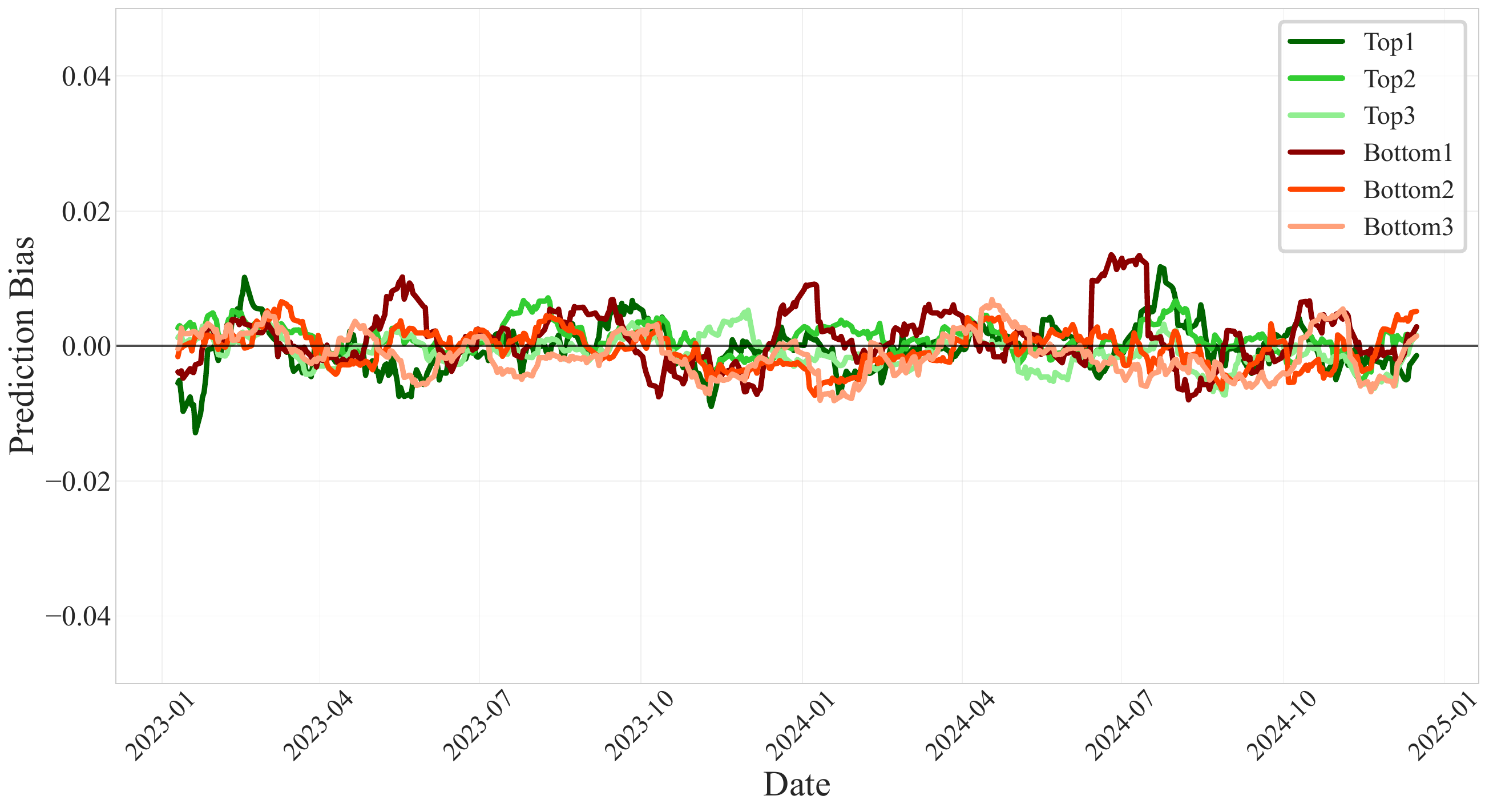}
    }
    \hfill
    \subfloat[$\mathbf{\alpha=1}$\label{fig:alpha1_subfig}]{%
        \includegraphics[width=0.49\textwidth]{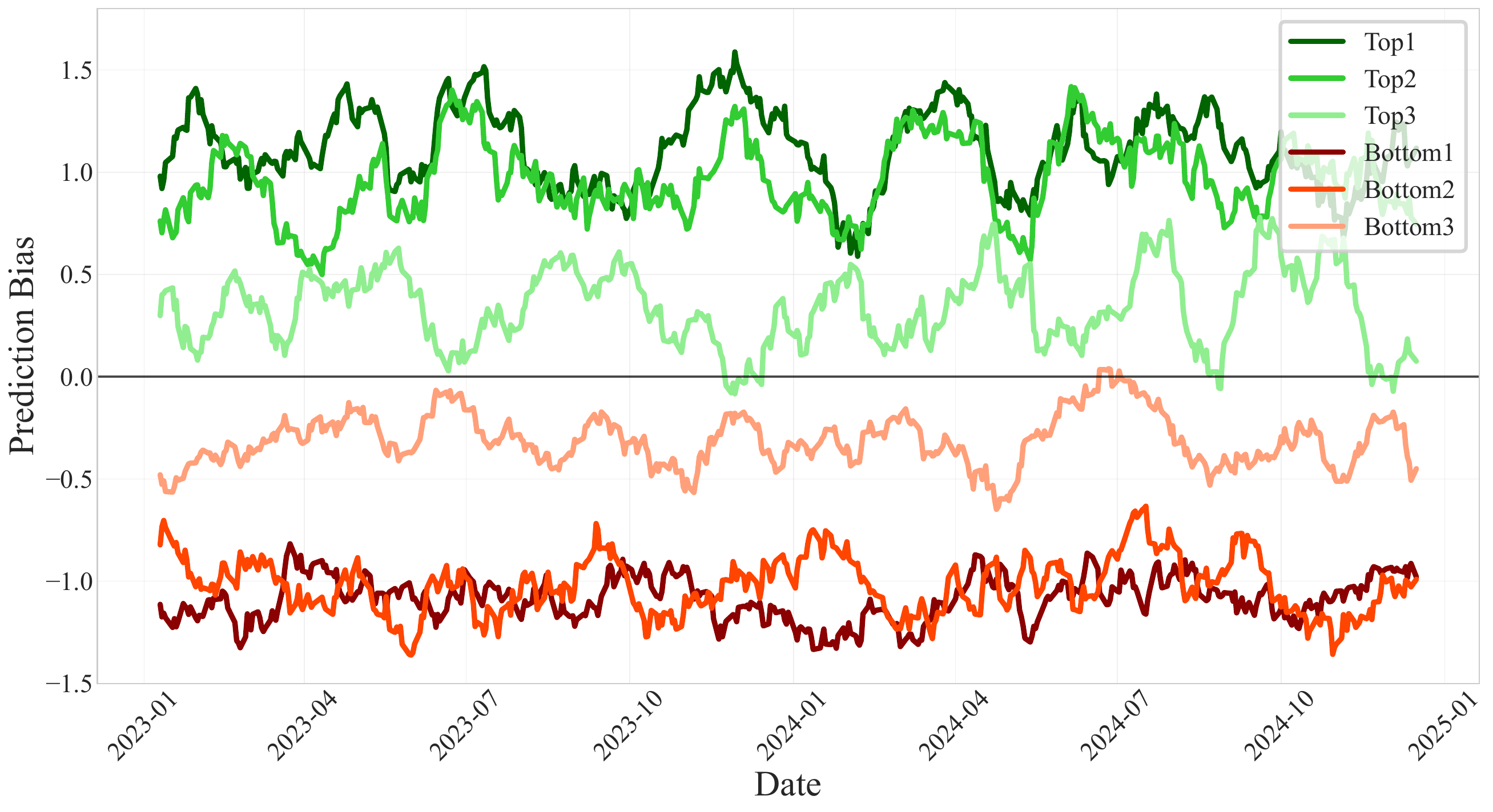}
    }
    \caption{Prediction bias patterns for portfolio assets under MSE loss ($\alpha=0$) versus MVO loss ($\alpha=1$). With MSE loss, prediction biases are randomly distributed across all assets regardless of portfolio inclusion. With MVO loss, prediction biases exhibit clear polarization: assets with high portfolio weights (Top) show positive bias while assets with low weights (Bottom) show negative bias, demonstrating how DFL induces strategic differentiation based on portfolio relevance.}
    \label{fig:bias_comparison_subfig}
\end{figure*}

\subsection{Prediction Bias}

\textbf{DFL and Prediction Accuracy.}  
Table~\ref{tab:Table1} shows that as the influence of the MVO loss increases, decision quality improves while MSE performance deteriorates. This phenomenon is commonly observed when applying DFL and has been discussed in prior studies~\cite{elmachtoub2022smart, mandi2022decision}. In this section, we aim to investigate how this trade-off affects the model's prediction behavior and contributes to improved decision-making. Figure~\ref{fig:mse_regret} shows that increasing the strength of the MVO loss leads to a consistent rise in the MSE loss. This trend is particularly pronounced when the model is trained purely with the MVO loss, where the MSE reaches its maximum compared to other $\alpha$ configurations. This behavior was observed consistently across all experimental settings. To better understand this phenomenon, we analyze prediction bias directly, defined as the discrepancy between the predicted return $\hat{\mu}$ and the ground truth return $\mu^*$. A positive prediction bias indicates that the model overestimates the true return, whereas a negative bias reflects underestimation.

\noindent  
\textbf{Prediction Bias across Up/Down Assets.}  
To examine how DFL affects prediction bias, we classify assets into two groups, Up and Down, based on whether the model predicts their returns to be positive or negative. We then analyze how the average bias in each group changes with increasing DFL strength.  
As shown in Figure~\ref{fig:Figure3}, stronger MVO loss supervision results in increasing polarization of prediction bias. That is, the average positive bias in the Up group and the average negative bias in the Down group both become more extreme. This suggests that stronger DFL encourages the model to make more confident directional predictions. Similar to the MSE degradation trend, this polarization effect was consistently observed across all asset universes and experimental settings.

\noindent  
\textbf{Prediction Bias across In/Out Assets.}
To analyze how prediction bias influences the actual portfolio decision, we group assets into two categories: those selected in the portfolio (In group) and those excluded (Out group). This grouping is performed at each optimization instance, and the predicted returns of each group are aggregated over the test period. This allows us to assess how the model differentiates between assets it selects and those it does not.

Figure~\ref{fig:predicted_return_distributions} presents the distribution of predicted returns for the In and Out groups under different training settings. As the strength of DFL increases, the separation between the two distributions becomes more distinct. This indicates that the model increasingly assigns higher positive bias to assets included in the portfolio, and greater negative bias to those excluded. In effect, the model amplifies the distinction between selected and non-selected assets, guiding the optimization toward more decisive selections.  
We exclude the case of $\alpha=1$ from the figure, as the predicted returns become excessively polarized, making comparison on a unified scale infeasible.

Another interesting observation is that as $\lambda$ increases, the overall predicted return distributions become flatter. This is not a numerical artifact, but rather evidence that the model is adapting to the shifting objective of the downstream optimization. In the MVO framework, a higher $\lambda$ reflects greater risk aversion, meaning the optimizer becomes more sensitive to portfolio variance than to expected return. As a result, the absolute magnitude of predicted returns plays a smaller role in portfolio selection. The DFL model internalizes this dynamic via the backpropagated MVO loss. It learns that when $\lambda$ is high, making sharp or highly differentiated predictions yields diminishing benefits for decision quality, as the optimizer focuses primarily on low-risk, low-correlation assets regardless of their predicted returns.

\noindent  
\textbf{Prediction Bias across Portfolio Weights.} To further understand how prediction bias relates to asset importance in the final decision, we analyze how prediction bias varies with the average portfolio weight of each asset. Figures~\ref{fig:alpha0_subfig} and \ref{fig:alpha1_subfig} present the time series of prediction biases for $\lambda=1.0$ under $\alpha = 0$ and $\alpha = 1$, respectively. The figures track six representative assets comprising the top three assets with the highest average portfolio weights and the bottom three with the lowest weights over the test period.

This analysis provides strong evidence for the strategic bias mechanism induced by DFL. Under MSE loss shown in Figure~\ref{fig:alpha0_subfig}, prediction biases for all assets fluctuate randomly around zero, regardless of their portfolio weights. In clear contrast, under pure DFL training in Figure~\ref{fig:alpha1_subfig}, a distinct polarization emerges where top-weighted assets consistently exhibit strong positive biases ranging from 0.5 to 1.5, while bottom-weighted assets show persistent negative biases from -0.5 to -1.5. This systematic pattern remains stable throughout the entire test period, demonstrating that DFL fundamentally alters the prediction model to create strategic differentiation between assets based on their portfolio relevance. These findings confirm that DFL's superior portfolio performance stems from its ability to amplify predictions for selected assets while suppressing those for excluded assets.


\section{Conclusion}

In this study, we investigated how Decision Focused Learning (DFL) fundamentally reshapes asset return prediction models within the Mean Variance Optimization (MVO) framework. Unlike traditional approaches that minimize uniform prediction errors such as MSE, DFL integrates the downstream portfolio optimization objective directly into the training process, guiding prediction models to prioritize decision quality over raw prediction accuracy. Our empirical results revealed three key findings. First, as demonstrated in Table 1, DFL leads to extreme portfolio concentration, with models consistently utilizing only the minimum number of assets allowed by constraints, which stands in stark contrast to models trained with MSE that maintain diversification. Second, DFL systematically induces prediction biases aligned with portfolio decisions, where assets selected for inclusion (IN group) exhibit positive bias while excluded assets (OUT group) show negative bias. This polarization intensifies with increasing DFL weight $\alpha$, creating increasingly distinct distributions between selected and non-selected assets. Third, our analysis of Up/Down asset groups confirmed that DFL amplifies directional predictions, with positive return assets becoming increasingly overestimated and negative return assets increasingly underestimated.

Theoretically, we explained this behavior through gradient analysis of the unconstrained MVO problem. We revealed that the DFL gradient $\frac{\partial U}{\partial \hat{\mu}_i} = \frac{1}{2\lambda} \Sigma^{-1}_i (\mu^* - \hat{\mu})$ fundamentally transforms MSE-based learning by tilting prediction errors with the inverse covariance matrix $\Sigma^{-1}$. This tilting mechanism explicitly incorporates inter-asset correlations into the prediction process, enabling the model to learn how assets move together rather than treating each prediction independently. This mechanism fundamentally differs from training based on MSE, which treats each asset's prediction error independently. Our findings resolve a long standing paradox in DFL regarding why DFL achieves superior portfolio performance despite higher prediction errors. The strategic prediction biases induced by DFL are not errors to be corrected but rather features that enhance decision quality. For practitioners, this insight suggests that when predictions are intended for portfolio optimization, traditional accuracy metrics may be misleading indicators of model quality.

\section{Limitations and Future Work}

While our study provides valuable insights into DFL mechanisms, there are several interesting avenues for future research that could further enrich our understanding. First, while we demonstrated that DFL induces systematic prediction biases across asset groups, a more granular analysis of individual asset characteristics could provide additional insights. Understanding which specific assets are more susceptible to strategic bias under DFL would complement our group-level findings and potentially inform more refined model architectures. Second, our theoretical analysis focused on the unconstrained MVO problem, which allowed us to derive closed-form solutions and clear insights. Extending this framework to constrained optimization settings such as those with budget limits, turnover restrictions, or market impact considerations represents a natural progression that would enhance the practical relevance of our findings.

Additionally, our results suggest an intriguing possibility regarding whether simpler alternatives could achieve similar benefits. Specifically, whether a learning approach based on MSE augmented with a covariance-weighted loss function might capture some of DFL's advantages remains an open and interesting question. Exploring such connections could lead to computationally efficient hybrid methods that bridge traditional and decision-focused approaches. These directions for future work build upon our core contribution of revealing how DFL systematically reshapes predictions through the $\Sigma^{-1}$ mechanism, offering exciting opportunities to further advance the integration of machine learning and portfolio optimization.

\begin{acks}
This work was supported by the National Research Foundation of Korea (NRF) grant funded by the Korea government (MSIT) (No. NRF-2022R1I1A4069163) and the Institute of Information \& Communications Technology Planning \& Evaluation (IITP) grant funded by the Korea government (MSIT) (No. RS-2020-II201336, Artificial Intelligence Graduate School Program (UNIST)). This work was also supported by the Institute of Information \& Communications Technology Planning \& Evaluation (IITP) grant funded by the Korea government (MSIT) (No. RS-2022-00143911, AI Excellence Global Innovative Leader Education Program).
\end{acks}

\bibliographystyle{ACM-Reference-Format}
\bibliography{ref}

\end{document}